\documentclass[11pt]{article}
\usepackage{times}
\usepackage{epsfig}
\usepackage{fullpage}
\usepackage{url}
\usepackage{textcomp}

\usepackage{color}  

\newcommand{\junk}[1]{}

\begin{document}


\title{Barcoding-free BAC Pooling Enables Combinatorial\\Selective Sequencing of the Barley Gene Space}

\author{Stefano Lonardi$^{1*}$, Denisa Duma$^1$, Matthew Alpert$^1$, Francesca Cordero$^{5,6}$,\\
Marco Beccuti$^5$, Prasanna R. Bhat$^{2,3}$, Yonghui Wu$^{1,4}$, Gianfranco Ciardo$^1$,\\
Burair Alsaihati$^1$, Yaqin Ma$^2$, Steve Wanamaker$^2$, Josh Resnik$^2$, Timothy J. Close$^2$\\ \ \\
\small $^1$ Department of Computer Science and Engineering, University of California, Riverside, CA 92521, USA \\
\small $^2$ Department of Botany \& Plant Sciences, University of California, Riverside, CA 92521, USA \\
\small $^3$ Monsanto Research Centre, Bangalore, 560092, India \\
\small $^4$ Google, Inc., Mountain View, CA 94043, USA \\
\small $^5$ Department of Computer Science, Universit\`{a} di Torino, 10149 Torino, Italy \\
\small $^6$ Department of Clinical and Biological Science, Universit\`{a} di Torino, 10043 Torino, Italy \\
\small $^*$ Corresponding author. Email: \url{stelo@cs.ucr.edu}}

\date{}
\maketitle

\begin{abstract} 
We propose a new sequencing protocol that combines recent advances in
combinatorial pooling design and second-generation sequencing
technology to efficiently approach \emph{de novo} selective genome
sequencing.  We show that combinatorial pooling is a cost-effective
and practical alternative to exhaustive DNA barcoding when dealing
with hundreds or thousands of DNA samples, such as genome-tiling
gene-rich BAC clones.  The novelty of the protocol hinges on the
computational ability to efficiently compare hundreds of million of
short reads and assign them to the correct BAC clones so that the
assembly can be carried out clone-by-clone. Experimental results on
simulated data for the rice genome show that the deconvolution is
extremely accurate (99.57\% of the deconvoluted reads are assigned to
the correct BAC), and the resulting BAC assemblies have very high
quality (BACs are covered by contigs over about 77\% of their length,
on average). Experimental results on real data for a gene-rich subset of
the barley genome confirm that the deconvolution is accurate (almost
70\% of left/right pairs in paired-end reads are assigned to the same
BAC, despite being processed independently) and the BAC assemblies have
good quality (the average sum of all assembled contigs is about 88\%
of the estimated BAC length).

\bigskip

\noindent\textbf{Data availability:} Barley raw sequencing data for
one set of 2,197 MTP gene-enriched BACs can be obtained from NCBI
Sequence Read Archive (\url{http://www.ncbi.nlm.nih.gov/sra?term=(SRA047913)})


\bigskip


\end{abstract}

\vfill
\newpage


\section*{\textsf{Introduction}}

The second generation of DNA sequencing instruments currently on the
market is revolutionizing the way molecular biologists design and
carry out scientific investigations in genomics and genetics.
Illumina, ABI SOLiD, Helicos, and Ion Torrent sequencing instruments
produce billions of sequenced reads at a fraction of the cost of
Sanger-based technologies, but read lengths are 100-150 bases, much
shorter than Sanger reads of typically 700--900 bases. While the
number (and to a lesser extent the length) of reads keeps increasing
at each update of these instruments, the number of samples that can be
run has remained small (e.g., two sets of seven \emph{lanes} on the
Illumina HiSeq). Since the number of reads produced by the instrument
is essentially fixed, when DNA samples to be sequenced are relatively
``short'' (e.g., BAC clones) and the correspondence between reads and
their source has to be maintained, several samples must be
``multiplexed'' on the same lane to optimize the trade-off between
cost and sequencing depth. Multiplexing is traditionally achieved by
adding a DNA barcode to each sample in the form of an additional
(oligo) adapter, but this does not scale readily to thousands of
samples. Although it is theoretically possible to barcode such a
number of samples, the procedure becomes unfeasible as the number of
sample is in the hundreds: the task is tedious, time consuming,
error-prone, and relatively expensive. While the task could be carried
out in principle by robotic instruments, most facilities do not have
access to these devices. Another significant disadvantage of
exhaustive barcoding is called ``barcoding bias'' which results in
very strongly non-uniform distribution of reads for each barcoded
sample (see, e.g., \cite{Alon:2011dz,CPSS08}).

In this paper, we demonstrate that multiplexing can be achieved
without exhaustive barcoding.  We propose a protocol based on recent
advances in combinatorial pooling design.  Combinatorial pooling has
been used previously in the context of genome analysis, but this is
the first attempt to use it for \emph{de novo} genome sequencing.
Earlier works are CAPSS and PGI, where BACs are arranged on a 2D
matrix, each row and column of the grid constituting a pool that is
then sequenced
\cite{Cai01,PGI_02,CAPS-MAP03,Milosavljevic2005}. However, this simple
grid design is very vulnerable to noise and behaves
poorly when several objects are positive; it is also far from optimal
in terms of the number of pools it produces.  Later works have
combined pooling with second-generation sequencing technology
\cite{SnehitPrabhu072009,YanivErlich072009,Hajirasouliha:2008}.  The
domain of application of ``DNA Sudoku'' is the detection of microRNA
targets in \emph{Arabidopsis} and human genes
\cite{YanivErlich072009}, whereas the pooling strategies discussed in
\cite{SnehitPrabhu072009,Hajirasouliha:2008} are used for targeted
resequencing (i.e., when a reference genome is available).  To the
best of our knowledge, there is no prior work on the feasibility of
combinatorial pooling and second-generation sequencing technology for
\emph{de novo} genome sequencing.


In our approach to \emph{de novo} sequencing, subsets of non-redundant
genome-tiling BACs are chosen to form intersecting pools. Each pool is
then sequenced individually on a fraction of a flowcell via standard
multiplexing.  Due to the short length of a BAC (typically
$\approx$130~kb), cost-effective sequencing requires each BAC pool to
contain hundreds or thousands of BACs.  We show later in this report
that attempting to directly assemble short reads originating from a
mix of hundreds BACs is likely to produce low-quality assemblies, as
the assembler is unable to partition the reads to individual
BACs. Moreover, it would be impossible to trace subset of the contigs
to a specific BAC address. If instead reads could be assigned (or
\emph{deconvoluted}) to individual BACs, then the assembly could
proceed clone-by-clone.  The goal of assigning sequenced reads to
specific BACs can be achieved if one chooses a pooling strategy in
which each BAC is present in a carefully designed set of pools such
that the identity of each BAC is encoded within the pooling pattern
(rather than by its association with a particular barcode). By
transitivity, the identity of each read is similarly encoded within
the pattern of pools in which it occurs.  Reads that can be assigned
to a given BAC are collected in a set, which is then individually
assembled.

To demonstrate the efficacy and performance of our approach, we apply
the proposed sequencing protocol to two plant genomes, namely rice and
barley, using the same pooling design parameters.  For rice, we started
from a fingerprint-based physical map, identified BACs on a minimum
tiling path (MTP), pooled the MTP BACs according to a shifted
transversal design \cite{Thierry2006a}, generated reads \emph{in
silico} from the pools (introducing some sequencing errors), deconvoluted the
reads to BACs, and finally assembled the reads clone-by-clone. The
sequence of the rice genome is used as the ``ground truth'' to
evaluate the accuracy of our method.  The results of the simulation
show that only 18.5\% of the short reads do not deconvolute.
The deconvolution process is surprisingly accurate: 99.57\% of the
deconvoluted reads are assigned to the original BAC. Most of the
non-assignable reads are those that appear in almost every pool, i.e.,
highly repetitive reads. An additional advantage of our approach is
protection against these repetitive reads, which can hamper the
assembly. We show that the resulting BAC assemblies have very high
quality, with large contigs covering on average 77\% of the rice BAC
sequence.

For barley, we also start from a fingerprint-based physical map of
gene-enriched BACs, identify BACs on a minimum tiling path (MTP), then
pool subsets of MTP BACs according to a shifted transversal
design. However, for barley we work on the actual clones and generate
sequence \emph{de novo}.  We are currently in the process of
sequencing seven sets of BAC pools, for a total of 14,763 MTP BACs.
Here, we report results on one set of 91 pools representing 2,197
BACs. In barley, a slightly larger number of reads did not deconvolute
due to the higher repeat content and length of this genome: 71.3\% of
the reads were assigned to 1--3 BACs, for a total of about 87\% of the
bases. The deconvolution process is also quite accurate on barley:
almost 70\% of left/right pairs in paired-end reads are assigned to
the same BAC, despite being processed independently.
The assembly statistics for barley show a lower average N50 than rice, but
the the average sum of all assembled contigs is about 88\%
the estimated BAC length. An objective measure of quality for barley BAC
assemblies is harder to implement due to the
absence of the ``ground truth''. As a workaround we measure the degree to which EST
consensus sequences (or \emph{unigenes}) known to be located in these
BACs are represented in the assemblies. The analysis shows that only 10\% of the final BAC
assemblies miss the expected unigene.  For the remaining 90\% of the
assemblies which contain the expected unigenes, the average coverage
of those unigenes is about 90\% of their length.  Finally, we compare
barley BAC assemblies to (1) the assembly of each pool of 169 BACs
(before deconvolution), (2) the assembly of the whole set of 91 pools
containing a total of 2,197 BACs (before deconvolution) and (3) the
assembly of the whole barley genome via shotgun sequencing (31x coverage). The
comparison shows that our BAC-by-BAC protocol is likely to be the most effective
strategy to obtain the largest possible N50\footnote{N50 indicates the
minimum length of all contig/scaffolds that together
account for 50\% of the genome.} for barley.


\section*{\textsf{Results}}

\subsection*{\textsf{Protocol overview}}

\begin{figure}[!ht]
  \centerline{\includegraphics[width=0.6\linewidth]{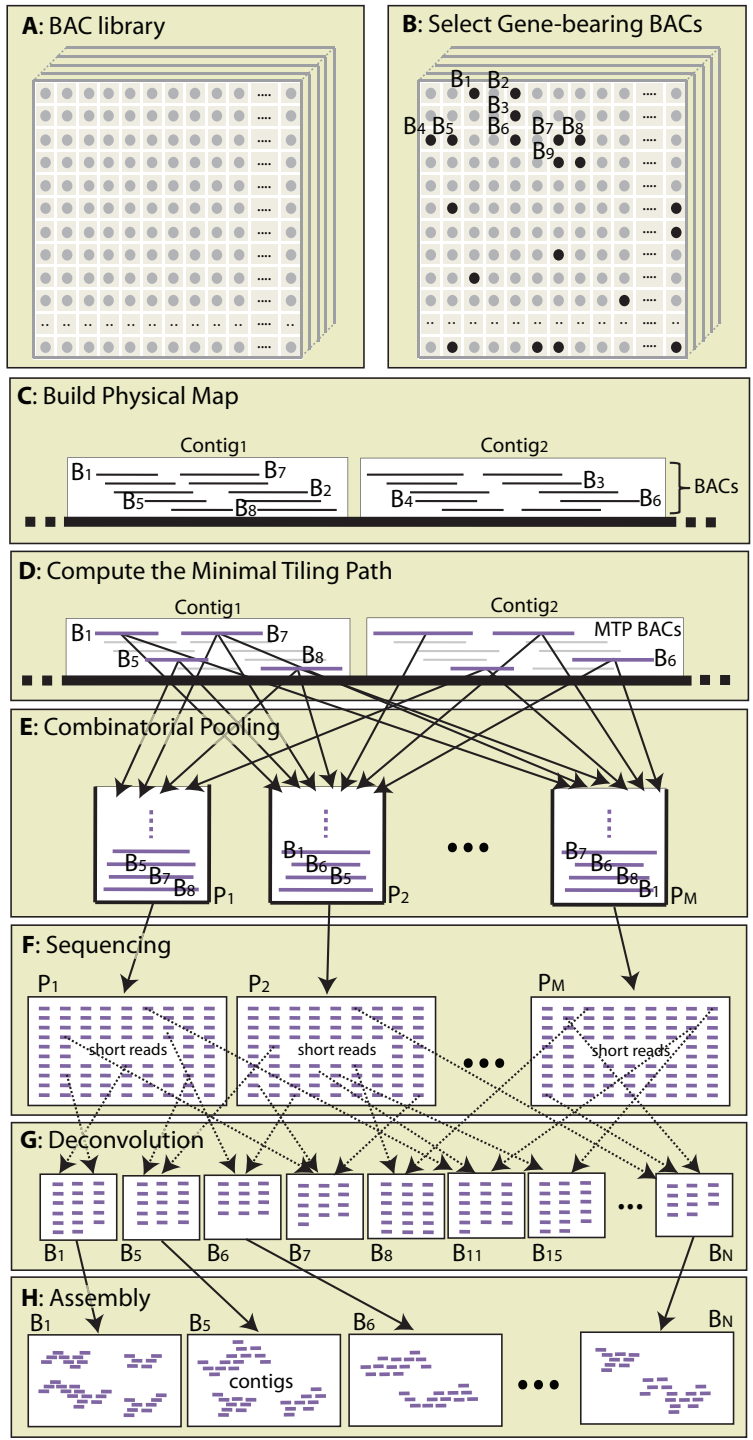}}
  \caption{Proposed sequencing protocol (see text for details).}
  \label{protocol}
\end{figure}

The main steps of our \emph{combinatorial sequencing} method are
summarized next and illustrated in Figure~\ref{protocol}. More details
can be found in the Methods section.

\begin{description}

\item[A.] Obtain a BAC library for the target organism

\item[B.] Select gene-enriched BACs from the library (optional)

\item[C.] Fingerprint BACs and build a physical map 

\item[D.] Select a minimum tiling path (MTP) from the physical map
\cite{FPC-MTP,BCL08}

\item[E.] Pool the MTP BACs according to the shifted transversal
design \cite{Thierry2006a} for an appropriate choice of $(P,L,\Gamma)$,
so that $P^{\Gamma+1} \geq N$, where $N$ is the number of BACs and
$\lfloor (L-1)/\Gamma \rfloor \geq 3$ (if the MTP was truly a set of
minimally overlapping clones, a two-decodable pooling would be
sufficient, but a three-decodable pooling gives additional protection
against errors)

\item[F.] Fragment the BACs in each pool, select for size, create a
 library for sequencing, sequence the DNA in each pool, trim reads
 based on quality scores, and remove \emph{E. coli} and vector
 contamination

\item[G.] Determine the \emph{signature} of each read via $k$-mer
analysis; assign reads to BACs by matching read signatures to BAC
signatures

\item[H.] Assemble reads assigned to each BAC into contigs using a
short-read assembler

\end{description}

Next, we report experimental results on simulated data on the genome
of \emph{Oryza sativa} (rice) and real sequencing data on the genome
of \emph{Hordeum vulgare L.} (barley).


\subsection*{\textsf{Simulation results on the rice genome}}

The physical map for \emph{Oryza sativa} was assembled from 22,474
BACs fingerprinted at AGCoL, and contained 1,937 contigs and 1,290
singletons. From this map, we selected only BACs whose sequence could
be uniquely mapped to the rice genome.  We computed an MTP of this
smaller map using our tool \textsc{FMTP} \cite{BCL08}.  The resulting
MTP contained 3,827 BACs with an average length of $\approx 150$~kb,
and spanned 91\% of the rice genome (which is $\approx 390$~Mb).

We pooled \emph{in silico} a subset of 2,197 BACs from the set above
according to the shifted transversal design \cite{Thierry2006a}.
Taking into consideration the format of the standard 96-well plate and
the need for a 3-decodable pooling design for minimal tiling path
BACs, we chose parameters $P = 13$, $L = 7$ and $\Gamma = 2$, so that
$P^{\Gamma+1} = 2{,}197$ and $\lfloor (L-1)/\Gamma \rfloor = 3$.  Each of
the $L = 7$ layers consisted of $P = 13$ pools, for a total of 91 BAC
pools, which left some space for a few control DNA samples on the
96-well plate. In this pooling design, each BAC is contained in
exactly $L = 7$ pools and each pool contains exactly $P^\Gamma = 169$
clones. The set of $L$ pools to which a BAC is assigned, is called the
\emph{BAC signature}. Any two pools can share at most $\Gamma=2$ BACs:
specifically, 57.9\% of the pool pairs have no BAC in common, 30.6\%
share one BAC, and 11.5\% share two.

The 91 resulting rice BAC pools were ``sequenced'' \emph{in silico} by
generating one million paired-end reads of 104 bases with an insert
size of 327 bases, and 1\% sequencing error distributed uniformly
along the read. A total of 208 million usable bases gave an expected $\approx
8$x depth of sequencing coverage for a BAC in a pool.  As each BAC
is present in seven pools, this is an expected $\approx 56$x
combined coverage.  We did not generate quality scores or artificially
introduced vector contamination, so cleaning step \textbf{F} was
irrelevant for these data.

The 91 read pools were processed for deconvolution using our tool
\textsc{HashFilter}. For each read in the pool set, this tool first
computes the occurrences of all distinct $k$-mers ($k=26$ in our
experiments) and stores them in a hash table.  Then it scans all the
reads again, this time fetching the set of pools that contain each
constitutive $k$-mer of a read, i.e., the \emph{$k$-mer signature}.
\textsc{HashFilter} compares each $k$-mer signature against the set of 2,197 BAC
signatures: if a match exists, that signature is declared
\emph{valid}.  Given the set of valid $k$-mer signatures for a read,
\textsc{HashFilter} finally determines the BACs to which the
read should be assigned (see Methods for more details).

The computation was not very time consuming, but required significant
amount of memory. On the rice dataset, the construction of the hash
table required about 120~GB of RAM and 164 minutes running on one core
of a Dell PowerEdge T710 server (dual Intel Xeon X5660 2.8Ghz, 12
cores, 144~Gb RAM). For the deconvolution phase, \textsc{HashFilter}
took 33 minutes running on 10 cores; sorting the reads into 2,197
files took 22 minutes (one core).

Figure~\ref{fig:read_distribution_on_rice}-TOP illustrates the
distribution of signature sizes for all the distinct $k$-mers in the
rice dataset. Observe that the distribution has clear peaks around
$L=7$, around the interval $[2L-\Gamma,2L]=[12,14]$ and the interval
$[3L-2\Gamma,3L]=[17,21]$. These peaks correspond to $k$-mers
originating from one, two, and three overlapping BACs, respectively.
We also have a rather large number of $k$-mers appearing in 1--5
pools. Observe that if the depth of sequencing coverage were high
enough and there were no sequencing errors, the minimum number of
occurrences for a $k$-mer would be $L=7$. For a $k$-mer to have fewer
than $L$ occurrences, sequencing errors must have occurred (assuming
the coverage to be sufficient).  However, these $k$-mers containing
sequencing errors are very likely to have an invalid signature, and
they will not be used to determine the signature of their
read. Figure~\ref{fig:read_distribution_on_rice}-BOTTOM shows the
distribution of signature sizes for all the reads in the rice dataset
built from valid $k$-mer signatures. The vast majority of reads now
have a signature size in the expected ranges, with the exception of
reads that appear in over 80 pools.  This latter set of reads cannot
be deconvoluted and is discarded, but this is a feature, not a
shortcoming, of our protocol: removing these ``ubiquitous'' reads
protects the assembly from highly repetitive reads, thus improves the
quality of BAC assemblies.

The set of reads with a signature of size 7, 12--14 or 17--21 that
could be deconvoluted was $\approx 81.5$\% of the total.
Table~\ref{tab:rice_pools} and Supplemental File~2 report the number
of reads assigned to one, two, or three BACs for each pool. Since we
knew the BAC from which each read was generated, the accuracy of the
deconvolution could be objectively determined. For rice, $99.57$\% of
the deconvoluted reads were assigned to either the correct BAC or to a
BAC overlapping the correct BAC (see Table~\ref{tab:rice_pools} and
Supplemental File~3).  After deconvolution, the average depth of
sequencing coverage for each BAC was $\approx 87$x, about 50\% higher
than the expected $56$x.  Even if we are losing about $18.5$\% of the
reads due to their invalid signatures, reads that can be deconvoluted
are frequently assigned to multiple BACs, thereby amplifying the
sequencing depth. Part of this inflation can be attributed to the
overlap between BACs in the MTP.

In the final step of the protocol, we independently assembled the set
of reads assigned to each BAC. We carried out this step with
\textsc{Velvet} \cite{Velvet08} for each of the 2,197 BACs, for a
variety of choices of $k$-mer size (hash length).  We also tested
\textsc{SOAPdenovo} \cite{Li:2010p1506} and \textsc{Abyss}
\cite{Simpson:2009}, but it was not obvious whether any of these
brought any improvement in the assemblies (data not shown). For
\textsc{Velvet}, we decided to report only the assembly that maximized
the N50. This is an arbitrary choice that does not guarantee the
``best'' overall assembly.  Supplemental File~3 reports all the
experimental results (see Supplemental Text for a description of each
column in the spreadsheet). If we average assembly statistics over all
the 2,197 BACs, the percentage of reads used in the assembly was
82.3\%, the average number of contigs was 41, the average N50 was
47,551~bp (31.4\% of the average BAC length), the average largest
contig was 57,258~bp (37.8\% of the average BAC length), the
average sum of all contig sizes was 137,050~bp (90.7\% of the
average BAC length).  The N50 is very high, and so is the percentage
of reads used by the assembler.  While these numbers already indicate
high quality assemblies, we wanted to determine whether BACs were
correctly assembled.  To do so, we \textsc{Blast}-ed the BAC contigs
against the rice genome and verified that \textsc{Velvet} actually
reconstructed the portion of the genome corresponding to the original
BAC.  For all 2,197 BAC assemblies, we recorded the fraction of each
original BAC covered by at least one contig, as well as the number and
average length of gaps and overlaps in the assemblies. Supplemental
File~3 (columns T--Y) shows these results. Considering these
statistics over all the 2,197 BACs, the average BAC coverage was
76.8\%, the average gap size was 263~bp, the average number of gaps
was 138, the average overlap size was 107~bp, and the average number
of overlaps was 75.


To establish a ``baseline'' for these assembly statistics, we considered
the most optimistic scenario of a ``perfect deconvolution'', which entails using the provenance
annotation of each read to assign it back to the correct BAC with
100\% accuracy.  Supplemental~File~4 reports the same statistics for
all 2,197 BAC assemblies in this best-case scenario.  If we compute the average
over all the 2,197 BACs, the average fraction of the reads used by
\textsc{Velvet} was 82.7\% and the average N50 was 132,865~bp (88\% of
the average BAC length). The \textsc{Blast} statistics showed an
average BAC coverage of 96.3\%, an average gap size of 52~bp, an
average number of gaps of 97, an average overlap size of 29~bp, and an
average number of overlaps of 54. Observe that, while the BAC coverage is about 20\% higher,
most of the other assembly statistics are comparable with the devolution via \textsc{HashFilter}.



  
\subsection*{\textsf{The gene-space of barley}}

Barley's diploid genome size is estimated at $\approx$ 5,300~Mb and it
composed of at least 80\% highly repetitive DNA, predominantly LTR
retrotransposons \cite{Wicker05}. The number of genes in barley is
uncertain; estimates range from 35,000 to 60,000.  Due to its size and
repeat content, a shotgun approach for \emph{de novo}
second-generation sequencing would require a very high depth of
sequencing, a mix of long insert paired-end reads of various length,
and the longest possible reads. Our protocol allows us to
tackle the assembly problem BAC-by-BAC, thus significantly reducing
its complexity and increasing the fidelity of the resulting
assemblies.

We started with a 6.3x genome equivalent barley BAC library created at
Clemson University Genomics Institute which contains 313,344 BACs with
an average insert size of 106~kb \cite{YTWFKKBMWW00}. Nearly 84,000
gene-enriched BACs were identified, mainly by the overgo probing
method \cite{KavithaNAR07} and [unpublished, 2011]. Gene-enriched BACs were fingerprinted at
University of California, Davis using high-information-content
fingerprinting \cite{Ding01,Luo03}. From the fingerprinting of
gene-bearing BACs, we produced a physical map \cite{BCL07,FPC-V4.7}
and derived a minimal tiling path of about 15,000 clones \cite{BCL08}.
Seven sets of $N =$~2,197 clones were chosen to be pooled according to
the shifted transversal design \cite{Thierry2006a}, which we
internally call HV3, HV4, \dots, HV9 (HV1 and HV2 were pilot
experiments). We used the same pooling parameters discussed in the
previous section ($P = 13$, $L = 7$ and $\Gamma = 2$). As a
consequence we had $P^{\Gamma+1} = N$ and decodability of $\lfloor
(L-1)/\Gamma \rfloor = 3$. Recall that in this pooling design, each
BAC is contained in exactly $L = 7$ pools and each pool contains
exactly $P^\Gamma = 169$ clones. Any two pools can share at most
$\Gamma=2$ BACs.

Here we are reporting on the HV5 set containing $91$ pools from a
total of 2,197 MTP gene-rich barley BAC clones.  Given the estimated
129.5~kb size of a BAC in the barley MTP (see section ``Clone-by-clone
Assembly'' for a discussion of the MTP BAC size estimate), the total
complexity of each pool of 169~BACs can be estimated at $\approx
22$~Mb.  As each BAC is replicated in seven pools, the total
complexity of the 2,197 BACs in HV5 is $\approx 286$~Mb.  To take
advantage of the high density of sequencing of the Illumina HiSeq2000,
we multiplexed thirteen pools on each lane using custom multiplexing
adapters. The total 91 pools used seven lanes, or one entire flowcell
of the instrument.

After reads were sequenced and demultiplexed, we obtained an average
of 12.4 million 94-base paired-end reads per pool. Reads were
end-trimmed using quality scores and kept only if longer than 36
bases, then cleaned of \emph{E. coli} contamination and spurious
Illumina adapters. The percentage of \emph{E. coli} in this particular
set of BACs was rather high, averaging around 51\%.  An alternative
DNA purification method can lower this amount to 8-10\% (see `Barley BAC pooling'
in Methods). Supplemental File~5 reports the number of reads and bases
after each step of the cleaning process.

The average number of usable paired-end reads after cleaning was about
5.5 million per pool with an average read length of 89 bases. The
distribution of the number of paired-end reads in the set of 91 pools
was between about 1M and 5.6M. Figure~1 in Supplemental Text
illustrates the number of single-end reads in each pool.  The total
number of paired-end reads for HV5 was about 250M, for a total of
about 44.8 billion usable bases. When compared to the 286~Mb
complexity of the sample, the average coverage (assuming a uniform
distribution) was $\approx$157x.

The 91 read pools in the barley HV5 dataset were processed using
\textsc{HashFilter}, and deconvoluted to one, two, or three
BACs. \textsc{HashFilter} built the hash table in about 340 minutes on
one core of a Dell PowerEdge T710 server and used about 43~Gb of
RAM. The deconvolution phase took 99 minutes on 10 cores, and the
sorting of reads into 2,197 files, one for each BAC, took 37 minutes
on one core. Due to the higher repeat content of the barley genome
compared to rice, \textsc{HashFilter} was able to deconvolute a
smaller fraction of the reads, about 71.3\% (see Table~1 in
Supplemental Text and Supplemental File~6). The total number of bases
was about 38.9 billion bases (about 87\% of the bases in HV5 before
deconvolution), which translated in an actual average coverage for
each BAC of about 137x (see Supplemental File~7, column~I).  While we
cannot objectively measure the accuracy of the deconvolution for
barley, six of the eight BACs that were assigned less than 20 reads
matched exactly the list of BACs that were noted as not growing during
the pooling carried out three years earlier (for a video of the
pooling see Supplemental File~1).

We carried out an analysis of deconvoluted paired-end reads, to
determine to what extent the left and the right mate agreed on their
BAC(s) assignment. \textsc{HashFilter} treats paired-end as two separate
single-end reads, which are deconvoluted independently. For each
paired-end read $r$, we collected in $L_r$ the set of BACs assigned to
the left mate, and in $R_r$ the set of BACs assigned to the right
mate.  Unless $L_r$ and $R_r$ were both empty, when $L_r \subseteq
R_r$ or $R_r \subseteq L_r$ we declared the paired-end read $r$ to be
\emph{concordant}. For barley, 68.7\% of the deconvoluted paired-end
reads were concordant, which indicates that the deconvolution was
quite accurate (see Supplemental File~6).  We note that about 22\% of
the paired-end reads in barley have one end for which the
corresponding BAC set is empty, probably due to sequencing errors or
repetitive content.  In this case, \textsc{HashFilter} does not
deconvolute the mate with the empty BAC set, and the other mate is
assigned to one or more BACs as a single-end read. We could have
modified \textsc{HashFilter} to exploit the paired-end association,
but that would have prevented us from carrying out this analysis.

We assembled each set of reads assigned to a BAC individually using
\textsc{Velvet} \cite{Velvet08} for a variety of choices of $k$-mer
size.  From the assemblies obtained for different choices of $k$, we
decided to report in Supplemental File~7 the assembly that maximized
the N50 (see Supplemental Text for a description of each column in the
spreadsheet).  If we average the assembly statistics over the 2,197
BACs, the number of reads used in the assemblies was 87.6\%,
indicating that \textsc{Velvet} took advantage of most of the data;
the average N50 was 7,210~bp (5.6\% of the average BAC length); the
average longest contig was 19,222~bp (14.9\% of the average BAC
length); the average sum of all the contigs in each assembly was
113,678~bp (87.8\% of the average BAC length).


  
\subsection*{\textsf{Barley Assembly Comparative Analysis and Validation via Illumina OPA}}

To understand the trade-offs between the number and the size
of the assembled contigs, the target size (e.g., BACs, set of BACs,
whole genome), and depth of sequencing coverage, we collected a set of
critical assembly statistics in Table~\ref{tab:assemblies}. The first
two rows contain average BAC assemblies statistics for rice data,
assuming perfect deconvolution or deconvolution via
\textsc{HashFilter}.

The average barley BAC assembly statistics are reported on the third
row, where reads were assigned to BACs via \textsc{HashFilter}, then
individually assembled with \textsc{Velvet}.  The next row represents
the average statistics obtained by assembling all the reads in each
pool of 169 BACs via \textsc{Velvet}, using the $k$-mer size that
maximized the N50 (see Supplemental File~7 for details). Recall that
each BAC is replicated in 7 distinct pools, so the depth of sequencing
coverage of one BAC in a pool is 1/7 of 180x, which is the coverage
before deconvolution. The fifth row reports the assembly of all the
reads in the 91 pools for HV5 using \textsc{SOAPdenovo}. Finally, the
last row reports the statistics of the whole shotgun assembly of the
barley genome using \textsc{SOAPdenovo} with $k=31$. The
whole shotgun sequencing of barley was carried out at several
locations: Ambry Genetics sequenced five (2$\times$77 bases)
paired-end lanes and four long-insert paired-end (LIPE) lanes (insert
size of 2, 3 and 5~kb); University of Minnesota (courtesy of Gary Muehlbauer) sequenced two
(2$\times$100 bases) paired-end lanes; University of California,
Riverside sequenced seven (2$\times$100 bases) paired-end lanes. The
number of usable paired-end bases after quality-based trimming was
159.31~Gb and 4.92~Gb of LIPE, for an overall 31x depth of
sequencing coverage of the 5.3~Gb barley genome.

Observe that as the target size increases from one BAC to the whole
genome, both the N50 and the number of reads used by the assembler
are monotonically decreasing, and so is sum of all contig sizes as a fraction
of the target size. This clearly indicates that
the effectiveness of the assembler decreases
as the complexity of the assembly problem increases,
which strongly advocates the use of a BAC-by-BAC
approach for the assembly of large, highly repetitive genomes.

Barley BAC assemblies were also compared against BAC-unigene lists
obtained using the Illumina GoldenGate oligonucleotide pool assay
(OPA) \cite{Fan:2006fk} developed for barley
\cite{TimBMCGenomics09}. We used the Illumina OPAs on the same seven
sets of barley pools described above (637 pools in total) and
determined which BAC clones were positive for two sets of 1,536 SNP
loci/unigenes (see Methods for details).

The GoldenGate assays allowed us to uniquely map a total of 1,849 
unique unigenes to BACs. Table~\ref{tab:OPA1} summarizes the results of unigene-BAC
BAC-unigene assignment broken down by chromosome and chromosome
arms.  The ratio of BACs to unigenes is 1.37, which
provides an estimate the amount of overlap among MTP clones.
BACs were anchored to a total of 333 unigenes mapped on
barley chromosome 5H, the maximum of any chromosome. Chromosome arm
5HL carries the maximum number of unigenes to which BACs were anchored
for a single arm at 253 unigenes. Supplemental File~9 contains all the solved BAC-unigene
relationships along with their chromosomal location.

Analysis of the assembly of the 2,197 barley BACs in the HV5 set was
carried out by using the results of the OPA as the ``ground
truth''. First, we extracted a total of 221 SNP loci/unigenes
(assembly B35) that were mapped to a total of 202 distinct BACs in
HV5.  We obtained the sequence of these 221 unigenes from
\textsc{Harvest} (\url{http://harvest.ucr.edu}) and \textsc{Blast}-ed
them against the HV5 BAC contigs. Out of 202 BACs that were expected
to contain those genes, only 20 BAC assemblies (10\%) missed entirely
the expected SNP loci/unigenes (see Supplemental File~7, columns
U--X).  For the other 90\% of the assemblies which contained the
expected unigenes, the average coverage of those unigenes was about
90\% of their length. This suggests that these BAC assemblies
contain the majority of the barley genes.



\section*{\textsf{Discussion}}

The challenges of \emph{de novo} sequence assembly originate from a
variety of issues, but two are the most prominent. First, sequencing
instruments are not 100\% accurate, and sequencing errors in the form
of substitutions, insertion, or deletions complicate the detection of
overlaps between reads. Second, large eukaryotic genomes contains many
highly repetitive elements. During the assembly process, all reads that
belong to those repetitive regions get \emph{over-compressed} and lead
to mis-assemblies.

To ameliorate the problems caused by repeats, two strategies can be
used, namely \emph{paired-end} and \emph{clone-by-clone} sequencing.
In paired-end sequencing, pairs of reads are obtained from both ends of
inserts of various sizes \cite{Roach95,Weber97}. Paired-end reads
resolve repeats during assembly simply by jumping across them
(abandoning the effort to fully resolve them) and disambiguating the
ordering of flanking unique regions.  Combined with shotgun
sequencing, this strategy has been successfully used to assemble
several complex genomes, including \emph{H. influenzae}
\cite{RDFleischmann07281995}, \emph{D. melanogaster}
\cite{Celera_Assembler}, \emph{H. sapiens} \cite{Venter01}, and
\emph{M. musculus} \cite{Waterston02} -- but with the caveat that the
resulting endpoint sequence is rarely 100\% complete.

In clone-by-clone sequencing, chunks of the genome (100--150~kb) are
cloned, typically in BACs, and then reads are obtained independently
from each clone \cite{Green2001}. By separating reads into sets that
represent individual BACs, sequences that are repetitive in the
context of the whole genome are more likely to have only a single copy in
each BAC; this greatly simplifies the assembly.  Hierarchical
sequencing was used to sequence several genomes including
\emph{S. cerevisiae} \cite{Oliver92,Yeast97}, \emph{C. elegans}
\cite{Celegans98}, \emph{A. thaliana} \cite{Ara00} and
\emph{H. sapiens} \cite{Lander01}.

The second generation of sequencing technologies based on flow cells
(e.g., Illumina, Helicos Heliscope and ABI SOLiD), has significantly
reduced the cost of sequencing, but the sequenced reads are much
shorter than Sanger reads. Shorter read length makes the problem of
\emph{de novo} genome assembly significantly harder. Although it has
been recently demonstrated that whole genome shotgun assembly from
short reads of a large eukaryotic genome (giant panda,
\emph{Ailuropoda melanoleura}) is possible \cite{Li:2010p1506}, the
contigs produced are relatively short, even considering the fact that
the sequencing depth was over 70x.

To the best of our knowledge, no clone-by-clone sequencing protocol
for second-generation instruments has been proposed so far.  We
believe that the major technical hurdle for a clone-by-clone approach
is the limitation of these instruments in handling hundreds or
thousand of BACs in a way that would allow reads to be assigned back
to their source. DNA barcoding can be used, but it does not scale well to
hundreds or thousands of samples, in part because an error rate of 0.1 to 1\%
confounds demultiplexing of incorrectly read barcode adapters.
Here, we have demonstrated an efficient alternative: instead of ligating barcodes to each BAC sample
before sequencing, we encode the ``signature'' of a BAC in the unique
set of pools to which it is assigned. By transitivity, reads
belonging to that BAC will also share the same signature.

Although our method is not entirely barcoding-free because we
multiplexed 13 BAC pools one the same lane of the sequencing
instrument, in principle it could be made completely free of DNA
barcodes by pooling a larger number of BACs and changing the pooling
parameters.  The decision to pool 2,197 BACs was made to be compatible
with the time required to manually create the pools in one day of work
for an average size lab.

Experimental results on simulated data for rice and actual sequencing
data for barley show that the clone-by-clone approach can be employed
with second-generation sequencing instruments.  Our method
deconvolutes reads to BAC with very high accuracy (99.57\% on rice),
and as a consequence the assemblies of the resulting BAC clones are of
high quality.  For the synthetic data (containing 1\% sequencing
errors) on the rice genome, we were able to reconstruct on average
77\% of the BACs content. On the barley data, the assembly
successfully reconstructed 90\% of the expected unigenes, with an
average coverage of the unigenes of about 90\%.
This amount of sequence will be adequate for most practical purposes such as map-based
cloning and nearby marker development for marker assisted breeding.

Combinatorial pooling provides an efficient approach
to clone-by-clone sequencing on second-generation instruments.
Clone-by-clone sequencing allows selectivity (e.g., gene-enriched
portion of a genome) and enables the distribution of the sequencing
work to multiple locations by partitioning the BACs to be sequenced.
It also decreases the sequencing and computational costs needed to
produce high quality assemblies, especially for large highly
repetitive genomes.  Combinatorial pooling has added benefits
which were not obvious before we started this project.  First, the
deconvolution process discards highly repetitive reads
without any prior knowledge; these repetitive reads would degrade the
assembly quality.  Second, pooling enables a very reliable detection
and correction of sequencing errors, a task currently under
development in our group.


\section*{\textsf{Methods}}

The steps in our \emph{combinatorial clone-by-clone sequencing} method
are illustrated in Figure~\ref{protocol} and described next is detail.


\subsection*{\textsf{Pooling (gene-rich) minimum-tiling-path BACs (Steps \textbf{A}-\textbf{E})}}

While our method can in general be applied to any set of clones that
cover a genome or a portion thereof, the protocol we are proposing
here for selective genome sequencing uses a physical map of
(gene-bearing) BACs to identify a set of minimally redundant
clones. The construction of a physical library and the selection of a
minimum tiling path are well-known procedures. More details can be
found in, e.g., \cite{Ding01,Luo03,FPC-V4.7,BCL07,BCL08} and
references therein.

Once the set of clones that need to be sequenced has been identified,
they must be pooled according to a scheme that allows the
deconvolution of the sequenced reads back to their corresponding
BACs. In Combinatorics, the design of a pooling method reduces to the
problem of constructing a \emph{disjunctive} matrix (see
\cite{Du1993}). Each row of the disjunctive matrix corresponds to a
BAC to be pooled and each column corresponds to a pool. Let us call
$w$ a subset of the rows (BAC clones) in the disjunctive matrix, and
let us define $u(w)$ as the set of pools that contain at least one BAC
in $w$. A design (or a matrix) is said to be $d$-\emph{decodable} if
$u(w_1) \not = u(w_2)$ when $w_1 \neq w_2$, $|w_1|\leq d$, and
$|w_2|\leq d$. The construction of $d$-decodable pooling designs has
been extensively studied \cite{Du1993}.  The popular 2D grid design is
simple to implement but cannot be used for the purposes of this work
because it is only one-decodable.

Recently, a new family of ``smart'' pooling methods has generated
considerable attention
\cite{Du06,Thierry2006a,Vermeirssen:2007,SnehitPrabhu072009,YanivErlich072009,Hajirasouliha:2008}. Among
these, we selected the \emph{shifted transversal} design
\cite{Thierry2006a} due to its ability of handling multiple positives
and its robustness to noise. The parameters of a shifted transversal
design pooling are defined by three integers $(P,L,\Gamma)$, where $P$
is a prime number, $L$ defines the number of layers, and $\Gamma$ is a
small integer. A \emph{layer} is a partition of BACs and consists of
exactly $P$ pools: the larger the number of layers, the higher is the
decodability. By construction the total number of pools is $P \times
L$.  If we set $\Gamma$ to be the smallest integer such that
$P^{\Gamma+1} \geq N$ where $N$ is the number of BACs that need to be
pooled, then the decodability of the design is $\lfloor (L-1)/\Gamma
\rfloor$.

An important property of this pooling design is that any two pools
only share at most $\Gamma$ BACs. By choosing a small value for
$\Gamma$ one can make pooling extremely robust to errors. In our
experiments, we use $\Gamma=2$, so that at least ten errors are needed
to mistakenly assign a read to the wrong BAC. In contrast, two errors
are sufficient to draw an erroneous conclusion with the 2D grid-design.



Barley BAC pools were obtained as follows.  \emph{Escherichia coli}
strain DH10B BAC cultures were grown individually in 96-well plates
covered by a porous membrane for 36 hr in 2YT medium with 0.05\%
glucose and 30 $\mu$g/ml chloramphenicol at 37\textcelsius\ in a
shaking incubator.  Following combinatorial pooling of 50 $\mu$l
aliquots from each of 169 BAC cultures, each of 91 collected pools
($\approx$8.3 ml each) was distributed into five 1.5 ml aliquots and
then centrifuged to create cell pellets. The pellets were frozen and
then used for extraction of BAC DNA using Qiagen plasmid DNA isolation
reagents.  Each BAC pool DNA sample was then dissolved in 45 $\mu$l of
TE buffer, and the five samples combined for a total of $\approx$225
$\mu$l at an estimated final concentration of 20 ng/$\mu$l. For
gene-BAC assignment using the Golden Gate assays, a total of 10 $\mu$l
($\approx$200 ng) of this DNA was then digested for 1 hour at
37\textcelsius\ by using 2 units of \emph{Not}I enzyme with 100
$\mu$g/ml BSA in a volume of 100 $\mu$l. The \emph{Not}I enzyme was
then heat inactivated at 65\textcelsius\ for 20 min.

BAC DNAs were prepared using a procedure that yields on average 65\%
BAC DNA and 35\% \emph{E. coli} DNA. Although these BAC DNAs performed
well for SNP locus detection in the GoldenGate assay, we were unaware
of the extent of \emph{E. coli} in the samples until we began BAC pool
sequencing, after all BAC pool DNAs had been prepared. Attempts were
made to remove \emph{E. coli} DNA from the BAC DNA samples through
selective digestion by using exonucleases, and to reduce highly
repetitive DNA using a denaturation/renaturation and double strand
nuclease method. These procedures provided little or no reduction of
the proportion of \emph{E. coli} DNA in the samples. A cost-benefit
analysis determined that the cost of replacing all of the BAC pools by
applying an alternative BAC DNA purification procedure yielding an
average of 94\% BAC DNA and 6\% \emph{E. coli} DNA would be no more
advantageous than simply repeating the sequencing of samples for which
more DNA sequence information was needed to support the
sequence-to-BAC deconvolution.

A video showing 44~seconds of the pooling process is available as
Supplemental File~1.


\subsection*{\textsf{Sequencing and Processing Paired-end Reads (step \textbf{F})}}

Sequencing of the barley BAC pools was carried out on an Illumina HiSeq~2000 at UC Riverside.
Paired-end reads from each pool were quality-trimmed
using a sliding window and a minimum Phred quality of 23. Next,
Illumina PCR adapters were removed with \textsc{Far} (Flexible Adapter
Remover, can be obtained from \url{http://sourceforge.net/projects/theflexibleadap/}), and discarded either if shorter than 36 bases or if
containing any `N'. Finally, reads were cleaned of \emph{E. coli}
(DH10B) and vector contamination (pBeloBAC11) using \textsc{BWA}
\cite{bwa09} and additional scripts.

According to our simulations, the depth of sequencing coverage of each
BAC after deconvolution is required to be at least 50x to obtain good
BAC assemblies.  The parameters of the pooling design should be chosen
so that the coverage pre-deconvolution is at least 150x-200x, to
compensate for non-uniformity in the molar concentrations of
individual BACs within each pool, BAC vector and \emph{E. coli}
contamination, and loss of reads due to the deconvolution
process.


\subsection*{\textsf{Deconvoluting Paired-end Reads to BACs (step \textbf{G})}}

\begin{figure}[!ht]
  \centerline{\includegraphics[width=0.99\linewidth]{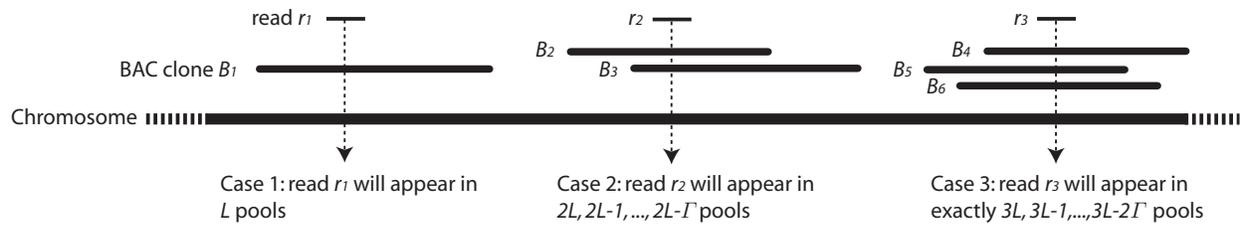}}
  \caption{An illustration of the three cases we are dealing with
   during the deconvolution process (clones belong to a MTP).}
  \label{example}
\end{figure}

\begin{figure}[!ht]
  \centerline{\includegraphics[width=0.85\linewidth]{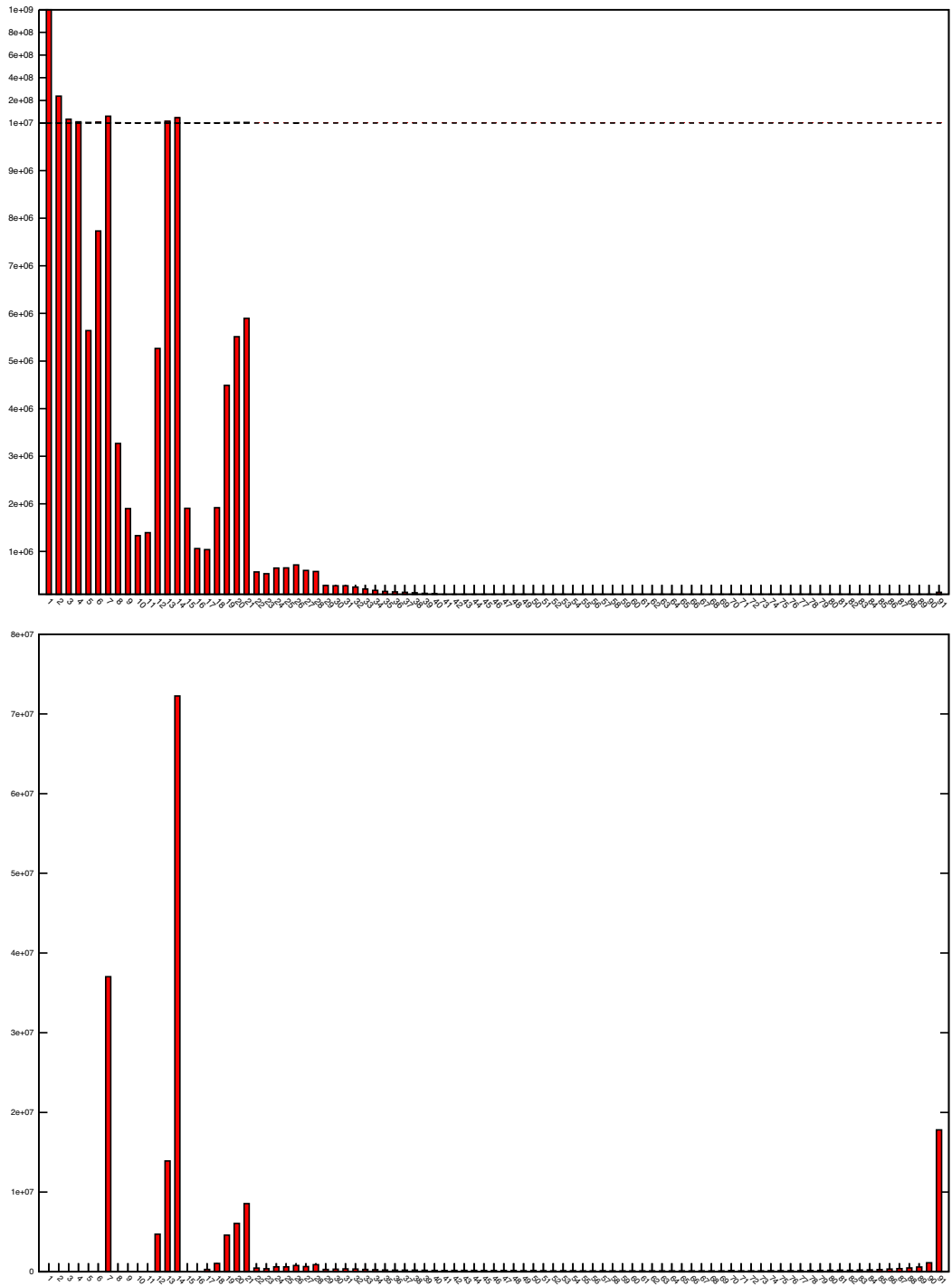}}
  \caption{Frequency distribution for the signatures of all distinct 26-mers (TOP) and all the reads (BOTTOM) in the 91 pools of synthetic sequencing data for rice; the x-axis represents the size of the signature, the y-axis is the frequency.}
  \label{fig:read_distribution_on_rice}
\end{figure}

To understand how deconvolution is achieved, let us make for a moment
the simplifying assumption that clones in the MTP do not
overlap. i.e., that the MTP BACs form a non-redundant tiling for the
genome under study, or a fraction thereof.  Let us pool the MTP BACs
according to a shifted transversal design with $L$ layers and obtain a
set of reads from them. Now, consider a read $r$ occurring only once
in the portion of the genome covered by the BACs. If there are no
sequencing errors and depth of sequencing is sufficient, $r$ will
appear in the sequenced output of exactly $L$ pools (see
Figure~\ref{example}, case 1). To determine the BAC to which a read
$r$ should be assigned, search for a BAC signature that matches the
list of positive pools for $r$.

For the most realistic scenario where at most $d$ MTP clones overlap,
the pooling must be at least $d$-decodable for the deconvolution to
work. We expect each non-repetitive read to belong to at most two BACs
if the MTP has been computed perfectly, or rarely three
BACs when considering imperfections, so we set $d=3$. When a read
belongs to the overlap between two clones (again assuming no
sequencing errors), it will appear in the sequenced output for $2L,
2L-1, \dots, 2L-\Gamma$ pools (see Figure~\ref{example}, case 2). The
case for three clones is analogous.

In general, the deconvolution method proceeds as follows. Recall that
in step \textbf{E} the number of pools is $M=P \times L$. Let us call
$R_i$ the set of reads obtained by sequencing pool $i$, for all $i \in
[1,M]$. For each set $R_i$, we first compute the frequency $count_i$ of
all its distinct $k$-mers. Specifically, for each $k$-mer $w \in R_i$,
$count_i(w)=c$ if $w$ or its reverse complement occurs exactly $c$
times in $R_i$.  These counts are stored in a hash table. For each
distinct $k$-mer $w$, the table stores a frequency vector of $M$
numbers, namely $[count_1(w), count_2(w), \dots, count_M(w)]$. Once
the table is built, we process each read as follows. Given a read $r$
in a pool, we fetch the frequency vectors for all its $k$-mers.
Recall that by construction each BAC is assigned to $L$ pools, thus
the \emph{signature} of a BAC is a set of $L$ numbers in the range
$[1,M]$. Due to our pooling design, two BAC signatures cannot share
more than $\Gamma$ numbers (see Theorem~I in
\cite{Thierry2006a}). Each $k$-mer signature is matched against the
BAC signatures, allowing for a small number of missing/extra pool
entries: if no good match exists, its frequency vector is discarded.
At the end of this process, the frequency vectors with a valid
signature are combined to form the \emph{signature} of read $r$. This
signature is matched again against the BAC signatures to determine the
BAC(s) to which it belongs. 

This algorithm is implemented in the tool \textsc{HashFilter} which
has been extensively tested under Linux platforms. The source code
and manual can be downloaded as Supplemental File~11.



\subsection*{\textsf{Clone-by-clone Assembly (step \textbf{H})}}

Once the reads were assigned to individual BACs, sets of single and
paired-end reads were assembled clone-by-clone using \textsc{Velvet}
\cite{Velvet08}. \textsc{Velvet} requires an expected coverage, which can
be computed using the amount of sequenced bases assigned to each BAC and
the estimated BAC size. For barley, BAC sizes were estimated from the number of
bands in the restriction fingerprinting data. First, we computed the average
number of bands in the 72,055 BACs fingerprinted at University of California, Davis
using high-information-content fingerprinting \cite{Ding01,Luo03} (see also
 \url{http://phymap.ucdavis.edu/barley/}).
Assuming that the average BAC length in this set was 106~kb, we computed
the multiplier to apply to the number of bands to obtain the estimated BAC length,
which turned out to be 1175 bases. We used that constant to obtain estimated sizes for all BAC in HV5
(see Supplemental File~7, column~F). Note that the average size of 129.5~kb is much larger than
the library average size of 106~kb; this indicates that the MTP selection favors larger BACs. 

We also tested \textsc{SOAPdenovo} \cite{Li:2010p1506} and \textsc{Abyss} \cite{Simpson:2009} on
simulated data (data not shown).  We evaluated the assembly for
several choices of the $k$-mer (hash) size, but only reported the
assembly that maximized the N50.  We recorded the number of contigs,
their N50/median/max/sum statistics, and the number of reads used in
the assembly.


For rice assemblies, we \textsc{Blast}-ed the BAC contigs to the rice
genome.  We computed the fraction of the original (source) BAC covered
by at least one contig, and the number of gaps and overlaps in the
assembly. The parameters used for \textsc{Blast} are reported in the
Supplemental Text.

For barley BAC assemblies, we carried out a validation based on the
known BAC-unigene associations from the Illumina GoldenGate assay
described in the next section. The validation involved
\textsc{Blast}-ing \#35 unigenes (Harvest:Barley assembly \#35
unigenes, \url{http://harvest.ucr.edu}) against the BAC assemblies.
To reduce spurious hits, we applied three filters.  First, we masked
highly repetitive regions by computing the frequency of all distinct
26-mers in the cleaned/trimmed HV5 data, then masking any $26$-mers
that occurred at least 11,000 times from the assembled contigs by
replacing the occurrences of those $k$-mers with Xs.  Second, we did
not consider a hit when a unigene was covered less than 50\% of its
length. Third, we excluded from the hit count any unigene that hit
more than ten individual BACs overall. We recorded the number
of unigenes hitting a BAC, and compared them with the expected
unigenes according to the Illumina assay.


\subsection*{\textsf{Barley GoldenGate oligonucleotide pool assay}}

Samples for the GoldenGate assay were prepared by combining 5 $\mu$l
of \emph{Not}I-digested BAC pool DNA ($\approx$10 ng) with 4 $\mu$l of
sonicated \emph{E. coli} DNA pre-dialyzed into TE buffer at a
concentration of 500 ng/$\mu$l (2000 ng) and 16 $\mu$l of TE
buffer. The final volume of each sample was thus 25 $\mu$l, composed
of $\approx$0.4 ng/$\mu$l of digested BAC pool DNA and 80 ng/$\mu$l of
additional \emph{E. coli} DNA. These DNA samples were provided to Joe
DeYoung at the University of California, Los Angeles, California, or
to Shiaoman Chao at the US Department of Agriculture genotyping
facility in Fargo, North Dakota. The DNA concentrations were then
readjusted to 50 ng/$\mu$l and a total of 5 $\mu$l of each DNA sample
was used for each GoldenGate assay.

Each Illumina GoldenGate oligonucleotide pool assay (OPA)
allows interrogation of a DNA sample for the presence
of 1536 SNP loci.  In \cite{TimBMCGenomics09}, five OPAs were designed
from approximately 22,000 SNPs from EST and PCR amplicon sequence
alignments.  Details of the development of three test phase (POPA1,
POPA2, and POPA3) and two production scale (BOPA1 and BOPA2) can be
found in \cite{TimBMCGenomics09}.

We genotyped the barley BAC pools described in Section ``The gene
space of barley'' on BOPA1 and BOPA2. Supplemental File~10 shows which
BOPA was applied to which set of barley BACs. The output from Illumina
GoldenGate assay was first converted to binary data by visual
inspection of the theta/R space in BeadStudio. A positive reading
meant that the SNP locus (and its corresponding unigene) is present in
at least one BAC within the pool (refer to Figure~2 in Supplemental Text for
an example).

Given the genotyping data for all unigene-pool pairs, we designed an
algorithm that computes the optimal assignment of unigenes to BACs so
that the number of errors is minimized.  For a particular unigene $g$
under consideration, let $O_g$ be the signature set of corresponding
positive pools. Let $S$ be an arbitrary set of BACs, where $1 \leq |S|
\leq 3$ and $U_S$ be the union of the pools that contain at least a
BAC clone in $S$. The number of errors $E_S$ associated with this
particular choice of $S$ is defined to be the number of extra
observations (equal to $|U_S \setminus O_g|$) plus the number of
missing observation (equal to $|O_g \setminus U_S|$). Among all
possible choices of $S$, we chose $S^*$ such that the value of
$E_{S*}$ is minimized.  When the number of errors associated with the
final solution was too large (say, more than 3), we declared that
unigene to be \emph{non-decodable}.

This procedure resulted in 1849 unigenes mapped to one,
two, or three BACs.  As a verification step, when a unigene was
mapped to more than one BAC, we verified that with a very low conflict frequency all those BACs belonged
to the same contig in the barley physical map \cite{BCL07,FPC-V4.7}.
Using the genetic map developed in \cite{TimBMCGenomics09,Munoz11} we were
also able to assign these unigene-anchored BACs to a barley genetic map position (Supplemental File 9).


\subsection*{\textsf{Data and Software Access}}

Barley raw sequencing data for the HV5 set can be obtained from NCBI
Sequence Read Archive (direct link \url{http://www.ncbi.nlm.nih.gov/sra?term=(SRA047913)}).
When  sequencing and  analysis are completed, we plan to release barley BAC assemblies for
each set of MTP BACs on \textsc{Harvest:Barley}
(\url{http://harvest.ucr.edu}) and \textsc{GenBank}
(\url{http://www.ncbi.nlm.nih.gov/genbank/}). The 31x shotgun genome assembly
of barley can be accessed via our \textsc{Blast} server 
hosted at the address \url{http://www.harvest-blast.org/}, by selecting ``Barley Genome''
from the database menu. This assembly will be made available on \textsc{Harvest:Barley}
(\url{http://harvest.ucr.edu}) and \textsc{GenBank}
(\url{http://www.ncbi.nlm.nih.gov/genbank/}).
The source code of \textsc{HashFilter} is available for download
as Supplemental File~11.


\section*{\textsf{Author Contributions}}

SL and TJC designed and supervised the project. SL wrote the initial
draft of the manuscript. TJC, PRB, SW, and JR produced the BAC pools
for barley. TJC supervised the collection of sequencing and genotyping
data for barley.  DD generated the synthetic data from the rice
genome, wrote a preliminary version of the tool to deconvolute reads
to BACs and evaluated the accuracy of the deconvolution.  TJC called the
Illumina OPA data to assign genes to BACs.  MA generated the
assemblies for rice and barley using \textsc{Velvet},
\textsc{SOAPdenovo}, and \textsc{Abyss} and wrote scripts to evaluate
their quality.  FC, MB, and GC designed and implemented the tool
\textsc{HashFilter} that computes the read signature and deconvolutes
of the reads.  YW wrote the tool to deconvolute SNP loci/unigenes to BAC
from the Illumina OPA data.  BA wrote a preliminary tool to compute
the all-pair prefix-suffix overlap using hash tables. YM prepared BAC DNA from
cell pellets and produced the sequencing libraries. SW wrote the
scripts to demultiplex and clean/trim the barley sequencing data.  All
authors read and approved the final manuscript.


\section*{\textsf{Acknowledgments}}

We thank M. C. Luo (UC Davis) for fingerprinting barley BACs and building an initial physical map, Serdar
Bozday (NIH) for building a compartmentalized physical map and selecting the MTP clones, Raymond D. Fenton
(UC Riverside) for assistance with BAC pooling and determining gene-BAC relationships using the Illumina
GoldenGate assay, Shiaoman Chao (USDA) and Joseph DeYoung (UC Los Angeles, Southern California Genotyping
Consortium) for providing Illumina GoldenGate services, and John Weger (UC Riverside Institute of
Integrative Genome Biology) for Illumina sequencing services. This research was supported in part by NSF
CAREER IIS-0447773, USDA 2009-65300-05645, NSF DBI-1062301, NSF 0321756, and USDA-CSREES-NRI
2006-55606-16722.


\pagestyle{empty}


\clearpage

\section*{Figures}

\vfill
\newpage

\vfill
\newpage

\vfill
\newpage

\begin{figure}[!ht]
  \centerline{\includegraphics[width=0.85\linewidth]{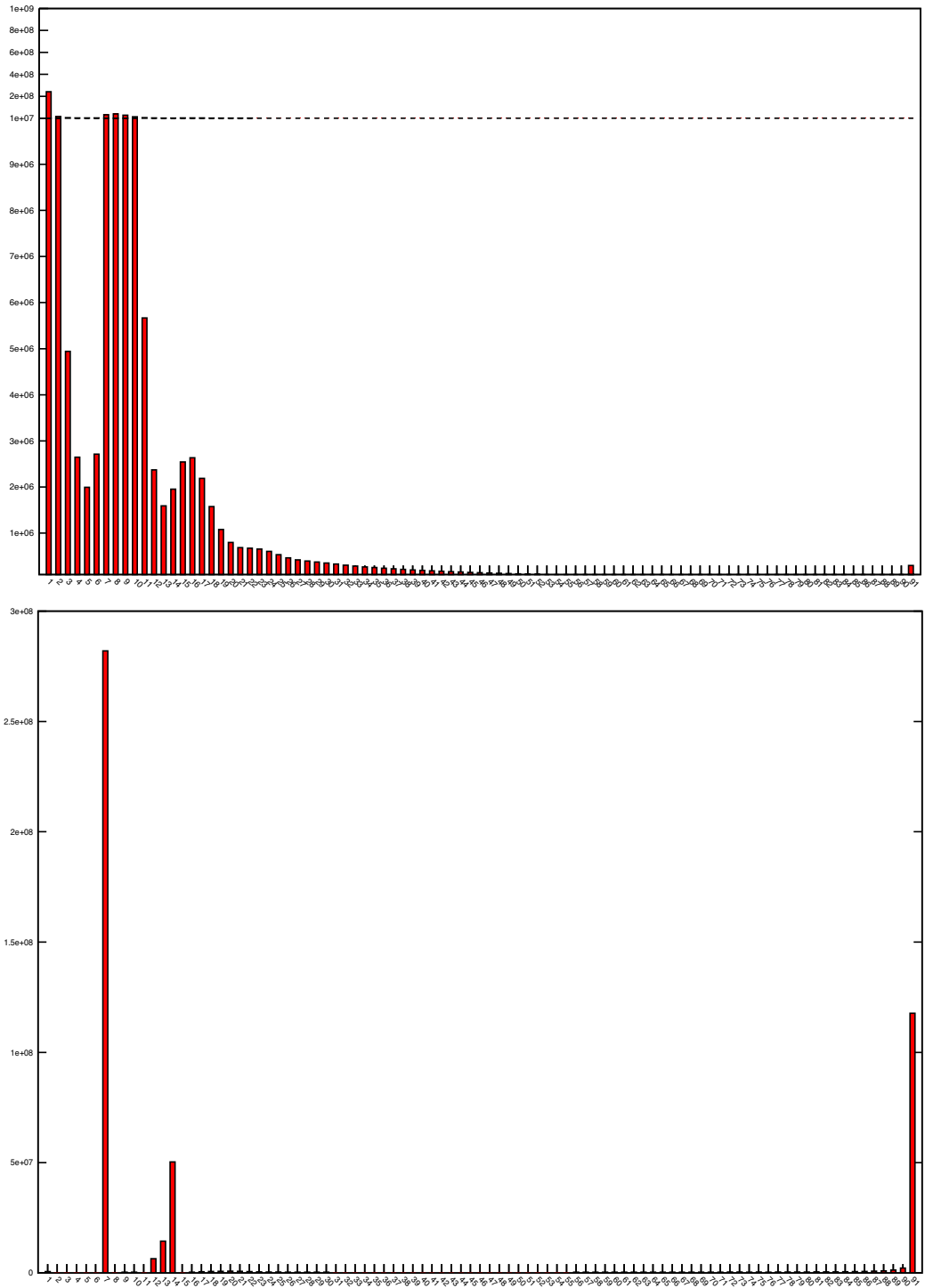}}
  \caption{Frequency distribution for the signatures of all distinct 26-mers (TOP) and all reads (BOTTOM) for a set of 91 pools of gene-enriched MTP BACs for barley (HV5); the x-axis represents the signature size, the y-axis is the frequency.}
  \label{fig:read_distribution_on_barley}
\end{figure}

\vfill
\newpage
\clearpage


\section*{Tables}

\begin{table}[ht]
\begin{center}
\begin{tabular}{ c | c c c | c }
  \hline
  Chromosome & Short arm & Long arm &  Uncertain & Total \\
  \hline
  1H         &      52 / \textcolor{red}{76}    &    140 / \textcolor{red}{201}   &        38 / \textcolor{red}{57} &   230 / \textcolor{red}{334} \\
  2H         &     114 / \textcolor{red}{181}   &    211 / \textcolor{red}{297}   &         2 / \textcolor{red}{5}  &   327 / \textcolor{red}{483} \\
  3H         &      80 / \textcolor{red}{119}   &    195 / \textcolor{red}{272}   &         0 / \textcolor{red}{0}  &   275 / \textcolor{red}{391} \\
  4H         &      74 / \textcolor{red}{103}   &    132 / \textcolor{red}{181}   &         1 / \textcolor{red}{1}  &   207 / \textcolor{red}{285} \\
  5H         &      68 / \textcolor{red}{94}    &    263 / \textcolor{red}{346}   &         2 / \textcolor{red}{3}  &   333 / \textcolor{red}{443} \\
  6H         &      77 / \textcolor{red}{116}   &    124 / \textcolor{red}{179}   &         0 / \textcolor{red}{0}  &   201 / \textcolor{red}{295} \\
  7H         &     146  / \textcolor{red}{207}  &    126 / \textcolor{red}{183}   &         0 / \textcolor{red}{0}  &   272 / \textcolor{red}{390} \\
  Unmapped   &           &          &        12{25} / \textcolor{red}{25} &    12 / \textcolor{red}{25} \\
  \hline
  Total      &     611 / \textcolor{red}{896}   &   1191 / \textcolor{red}{1659}   &        55 / \textcolor{red}{91} &  1857 / \textcolor{red}{2646} \\
  Unique     &           &          &           &  1849 / \textcolor{red}{2541}
\end{tabular} 
\end{center}
\caption{Chromosomal distribution of unigenes (assembly \#35) contained in BACs (black numbers), and BACs containing unigenes (red numbers), according to GoldenGate assays.}
\label{tab:OPA1}
\end{table}

\vfill
\newpage

\begin{table}[ht]
 \begin{center}
 \scriptsize{
 	\begin{tabular}{rrrrrr||rrrrrr}
 	\hline
 	\textbf{Pool} & \textbf{1 BAC} & \textbf{2 BACs} & \textbf{3 BACs} & \textbf{\%Total} & \textbf{\%Correct} & \textbf{Pool} & \textbf{1 BAC} & \textbf{2 BACs} & \textbf{3 BACs} & \textbf{\%Total} & \textbf{\%Correct}\\
 	\hline
 1 & 390,925 & 1,021,202 & 218,965 & 81.55\% & 99.57\% & 47 & 415,306 &   958,780 & 238,508 & 80.63\% & 99.53\% \\
 2 & 401,930 & 1,010,577 & 226,119 & 81.93\% & 99.45\% & 48 & 395,628 &   988,326 & 244,361 & 81.42\% & 99.39\% \\
 3 & 446,845 & 1,019,560 & 199,438 & 83.29\% & 99.59\% & 49 & 341,453 & 1,062,420 & 232,221 &  81.8\% & 99.55\% \\
 4 & 460,513 & 1,012,335 & 187,729 & 83.03\% & 99.59\% & 50 & 440,861 &   943,720 & 244,218 & 81.44\% & 99.55\% \\
 5 & 455,705 &   947,515 & 222,519 & 81.29\% & 99.52\% & 51 & 441,704 &   983,305 & 192,133 & 80.86\% & 99.55\% \\
 6 & 391,456 & 1,044,628 & 198,262 & 81.72\% & 99.63\% & 52 & 410,569 &   962,425 & 242,805 & 80.79\% & 99.62\% \\
 7 & 391,010 & 1,045,500 & 230,553 & 83.35\% & 99.62\% & 53 & 400,208 & 1,018,204 & 211,243 & 81.48\% & 99.53\% \\
 8 & 388,850 &   991,831 & 243,614 & 81.21\% & 99.58\% & 54 & 380,140 &   969,012 & 253,479 & 80.13\% & 99.49\% \\
 9 & 381,752 &   975,607 & 256,767 & 80.71\% & 99.58\% & 55 & 420,342 & 1,007,343 & 211,307 & 81.95\% & 99.56\% \\
10 & 346,968 & 1,045,075 & 243,428 & 81.77\% & 99.64\% & 56 & 449,944 &   979,593 & 215,811 & 82.27\% & 99.57\% \\
11 & 394,704 &   964,910 & 227,215 & 79.34\% & 99.57\% & 57 & 393,856 & 1,060,639 & 209,274 & 83.19\% & 99.55\% \\
12 & 420,363 &   936,500 & 222,287 & 78.96\% & 99.49\% & 58 & 368,063 & 1,062,521 & 224,716 & 82.77\% & 99.62\% \\
13 & 411,143 &   969,745 & 239,441 & 81.02\% &  99.6\% & 59 & 382,411 & 1,064,979 & 191,622 & 81.95\% & 99.56\% \\
14 & 386,831 & 1,028,001 & 226,478 & 82.07\% & 99.62\% & 60 & 394,017 &   992,574 & 235,947 & 81.13\% & 99.62\% \\
15 & 360,496 & 1,053,183 & 245,686 & 82.97\% & 99.59\% & 61 & 428,393 &   968,451 & 234,017 & 81.54\% & 99.62\% \\
16 & 413,108 & 1,031,143 & 193,165 & 81.87\% &  99.6\% & 62 & 511,416 &   934,536 & 204,130 &  82.5\% & 99.59\% \\
17 & 426,155 &   984,613 & 202,242 & 80.65\% & 99.54\% & 63 & 323,162 & 1,019,112 & 248,197 & 79.52\% & 99.41\% \\
18 & 425,161 &   972,202 & 229,742 & 81.36\% & 99.55\% & 64 & 447,481 &   936,733 & 230,762 & 80.75\% & 99.64\% \\
19 & 377,124 &   993,507 & 256,679 & 81.37\% & 99.62\% & 65 & 392,007 &   968,324 & 242,245 & 80.13\% & 99.58\% \\
20 & 392,747 & 1,012,836 & 216,100 & 81.08\% & 99.55\% & 66 & 346,148 & 1,021,575 & 231,422 & 79.96\% & 99.49\% \\
21 & 358,849 & 1,016,130 & 237,873 & 80.64\% & 99.49\% & 67 & 410,069 &   922,582 & 230,421 & 78.15\% & 99.38\% \\
22 & 438,686 &   998,197 & 214,066 & 82.55\% & 99.55\% & 68 & 432,649 &   952,708 & 224,847 & 80.51\% & 99.56\% \\
23 & 440,145 &   959,963 & 230,235 & 81.52\% & 99.61\% & 69 & 373,656 &   983,368 & 260,897 &  80.9\% & 99.61\% \\
24 & 470,767 &   970,915 & 207,429 & 82.46\% & 99.48\% & 70 & 399,624 & 1,040,903 & 203,423 &  82.2\% & 99.61\% \\
25 & 413,950 &   968,673 & 229,748 & 80.62\% & 99.57\% & 71 & 417,006 & 1,032,484 & 204,127 & 82.68\% & 99.59\% \\
26 & 380,879 &   993,435 & 225,879 & 80.01\% & 99.56\% & 72 & 430,118 & 1,002,137 & 212,012 & 82.21\% & 99.51\% \\
27 & 409,336 & 1,011,286 & 204,935 & 81.28\% & 99.61\% & 73 & 430,159 & 1,020,881 & 199,634 & 82.53\% & 99.66\% \\
28 & 413,659 &   970,618 & 230,663 & 80.75\% & 99.58\% & 74 & 389,350 &   983,376 & 246,105 & 80.94\% & 99.56\% \\
29 & 478,045 &   956,413 & 219,851 & 82.72\% & 99.63\% & 75 & 485,180 &   968,794 & 207,182 & 83.06\% & 99.58\% \\
30 & 437,710 &   958,018 & 239,748 & 81.77\% & 99.62\% & 76 & 427,825 &   999,602 & 229,448 & 82.84\% & 99.64\% \\
31 & 312,489 & 1,051,366 & 263,548 & 81.37\% & 99.59\% & 77 & 348,986 & 1,047,405 & 238,092 & 81.72\% & 99.61\% \\
32 & 399,797 & 1,001,191 & 223,652 & 81.23\% & 99.45\% & 78 & 394,006 & 1,008,216 & 227,993 & 81.51\% & 99.57\% \\
33 & 368,754 & 1,049,749 & 209,490 &  81.4\% & 99.51\% & 79 & 349,668 & 1,028,294 & 220,179 & 79.91\% & 99.51\% \\
34 & 394,542 & 1,029,862 & 193,241 & 80.88\% & 99.54\% & 80 & 422,887 &   975,647 & 231,153 & 81.48\% & 99.46\% \\
35 & 384,702 & 1,067,284 & 185,431 & 81.87\% & 99.51\% & 81 & 404,990 &   998,443 & 222,417 & 81.29\% & 99.55\% \\
36 & 381,991 &   988,185 & 228,927 & 79.96\% & 99.52\% & 82 & 429,214 &   985,137 & 219,765 & 81.71\% & 99.58\% \\
37 & 447,843 &   909,650 & 263,912 & 81.07\% & 99.52\% & 83 & 391,829 & 1,046,567 & 175,777 & 80.71\% & 99.56\% \\
38 & 453,436 &   996,709 & 210,853 & 83.05\% & 99.69\% & 84 & 418,626 &   999,325 & 229,379 & 82.37\% & 99.63\% \\
39 & 403,907 &   995,495 & 241,792 & 82.06\% & 99.57\% & 85 & 412,015 &   977,566 & 239,830 & 81.47\% & 99.62\% \\
40 & 383,803 & 1,010,100 & 230,899 & 81.24\% & 99.56\% & 86 & 415,146 &   973,912 & 220,785 & 80.49\% & 99.43\% \\
41 & 406,096 &   998,748 & 211,387 & 80.81\% & 99.53\% & 87 & 399,776 &   980,706 & 242,806 & 81.16\% & 99.57\% \\
42 & 457,064 & 1,002,075 & 188,545 & 82.38\% & 99.62\% & 88 & 372,785 & 1,026,350 & 238,468 & 81.88\% & 99.57\% \\
43 & 430,717 & 1,028,726 & 196,809 & 82.81\% &  99.6\% & 89 & 436,803 &   988,090 & 225,036 &  82.5\% & 99.63\% \\
44 & 380,624 & 1,031,742 & 210,096 & 81.12\% & 99.63\% & 90 & 385,299 & 1,018,026 & 238,231 & 82.08\% & 99.62\% \\
45 & 357,076 &   992,491 & 257,128 & 80.33\% & 99.58\% & 91 & 445,759 &   986,738 & 205,512 &  81.9\% & 99.59\% \\
46 & 426,502 & 1,023,330 & 224,662 & 83.72\% & 99.62\% & Avg      &   406,612 & 998,798 & 224,167 & 81.48\% & 99.57\% \\
 	\hline
 	\end{tabular}
 }
 \end{center}
\caption{Number of reads per pool deconvoluted to one, two, or three BACs; the percentage column 
reports the fraction of the total number of reads that were deconvoluted and
the total number of correct reads (rice synthetic data).}
\label{tab:rice_pools}
\end{table}

\vfill
\newpage


\begin{table}[ht]
  \begin{center}
    \begin{tabular}{ l r | r r r r }
      \hline
      \emph{Target}                                            & \emph{Size} (Mb) & \emph{Coverage}     & \emph{\% reads used}$^{c}$ & \emph{L50} (bp) & \emph{\% Sum} \\
      \hline
      Rice, 1 BAC (perfect deconvolution)$^{a}$                & 0.151   &  56x                & 82.7\%        & 132,865 & 98.7\% \\
      Rice, 1 BAC (\textsc{HashFilter} deconvolution)$^{a}$    & 0.151   &  87x                & 82.3\%        &  47,551 & 90.7\% \\
      \hline
      Barley, 1 BAC (\textsc{HashFilter} deconvolution)$^{a}$  & 0.129   &  137x               & 87.6\%        &   7,210 & 87.8\% \\
      Barley, 169 BACs$^{b}$ (no deconvolution)                &    20   &  25.7x              & 67.1\%        &   4,270 & 69.5\%  \\
      Barley, 2,197 BACs ($k=25$, no deconvolution)             &   250   &  180x               & 25.3\%        &   3,845 & 56.6\% \\
      Barley, whole genome ($k=31$)                            & 5,300   &  31x                & 13.3\%        &   2,857 & 30.5\% \\
      \hline
    \end{tabular} 
  \end{center}
  \caption{Summary of the statistics of the various assemblies obtained
    using \textsc{Velvet} (rows 1--4) and \textsc{SOAPDenovo} (rows 5--6); ``\% Sum'' is the
    the sum of all contig sizes as percentage of the target size;
    $^{a}$average over 2,197 assemblies; $^{b}$average over 91 assemblies;
    $^{c}$\textsc{Velvet} reports the number of reads used in the assembly but
    \textsc{SOAPDenovo} does not: for the last two assemblies, we used
    \textsc{Bowtie} to align reads to the assemblies (1 mismatch) }
  \label{tab:assemblies}
\end{table}

\clearpage


\small
\bibliography{t}
\bibliographystyle{apalike}

\normalsize

\end{document}